\documentclass[%
 reprint,
showpacs,preprintnumbers,
 amsmath,amssymb,
 aps,
]{revtex4-1}

\usepackage[dvipdfmx]{graphicx}
\usepackage{epstopdf}
\usepackage{dcolumn}
\usepackage{bm}
\usepackage{color}
\usepackage{ulem}

\begin{document}

\title{Temperature scaling in nonequilibrium relaxation in three-dimensional \\
Heisenberg model in the Swendsen-Wang and Metropolis algorithms}

\author{Yoshihiko Nonomura}
\email{nonomura.yoshihiko@nims.go.jp}
\affiliation{International Center for Materials Nanoarchitectonics, 
National Institute for Materials Science, Tsukuba, Ibaraki 305-0044, Japan} 

\author{Yusuke Tomita}
\email{ytomita@shibaura-it.ac.jp}
\affiliation{College of Engineering, Shibaura Institute of Technology, 
Saitama 337-8570, Japan} 

\begin{abstract}
Recently, the present authors proposed the nonequilibrium-to-equilibrium 
scaling (NE-ES) scheme for critical Monte Carlo relaxation process, which 
scales relaxation data in the whole simulation-time regions regardless 
of functional forms, namely both for the stretched-exponential critical 
relaxation in cluster algorithms and for the power-law critical relaxation 
in local-update algorithms. In the present study, we generalize this 
scheme to off-critical relaxation process, and scale relaxation data 
for various temperatures in the whole simulation-time regions. 
This is the first proposal of the off-critical scaling in cluster algorithms, 
which cannot be described by the dynamical finite-size scaling theory 
based on the power-law critical relaxation. As an example, we investigate 
the three-dimensional Heisenberg model previously analyzed with the NE-ES 
[Y.~Nonomura and Y.~Tomita, Phys.\ Rev.\ E {\bf 93}, 012101 (2016)] 
in the Swendsen-Wang and Metropolis algorithms. 
\end{abstract}


\maketitle

\section{Introduction}
The nonequilibrium relaxation (NER) method is one of the improved Monte 
Carlo schemes to study phase transitions against the critical slowing down. 
In general, basic formulation of the NER method is based on the critical 
relaxation, and off-critical behaviors are described by scaling analyses. 
In local-update algorithms, the critical relaxation is characterized by 
the power-law behavior of physical quantities, and the critical point is 
determined as the most probable point to exhibit such a behavior~\cite{NERrev}. 
This NER behavior is derived from the dynamical finite-size scaling 
(DFSS) theory~\cite{Suzuki76,Hohenberg77}, and the off-critical 
scaling behavior is also derived from it.

Recently, the present authors revealed that the critical NER 
behaviors in cluster algorithms~\cite{SW,Wolff} are described 
by the stretched-exponential simulation-time dependence in various 
classical spin systems~\cite{Nonomura14,Nonomura15,Nonomura16} 
and in a quantum phase transition~\cite{Nonomura20}. 
Although the critical point can be determined from such 
early-time relaxation behaviors, more precise estimation 
is possible from the nonequilibrium-to-equilibrium scaling 
(NE-ES)~\cite{Nonomura14,Nonomura16,Nonomura20}, 
which connects the early-time and equilibrium behaviors smoothly. 
In addition to these numerical findings, the present authors derived 
this relaxation formula phenomenologically in the Ising models 
in the Swendsen-Wang (SW) algorithm~\cite{Tomita18}. 

Although the DFSS is not defined in cluster algorithms, in the present 
article we generalize the NE-ES to the off-critical region and confirm 
this novel ``temperature scaling" in the three-dimensional (3D) 
Heisenberg model in the SW algorithm, which we analyzed precisely 
with the NE-ES~\cite{Nonomura16}. Here we also show that this new 
formalism is applicable even to local-update algorithms.

The outline of the present article is as follows: In Section II, we briefly 
summarize the model and  Monte Carlo method used in the present 
article, and review the NER method, the DFSS and the NE-ES. 
In Section III, we derive the temperature scaling in cluster and 
local-update algorithms, and compare the formula with the one 
obtained from the DFSS. In section IV, we numerically confirm 
the temperature scaling with the magnetic susceptibility in the 
3D Heisenberg model. As typical cluster and local-update 
algorithms, the SW and Metropolis ones are utilized. 
In the Metropolis algorithm, the conventional scaling 
analysis based on the DFSS is also made for comparison. 
In section V, these results are compared with each other 
and with the previous numerical results, and we propose 
a general framework to investigate critical phenomena 
efficiently by combining the present scheme and the NE-ES. 
The above descriptions are summarized in Section VI. 
In the appendix, similar analyses on the absolute value 
of magnetization are summarized. 

\section{Model and method}
In the present article, the 3D Heisenberg model on a cubic lattice 
described by the following Hamiltonian, 
\begin{equation}
\label{Hei-Ham}
{\cal H}=-J\sum_{\langle ij \rangle \in {\rm n.n.}}\vec{S}_{i} \cdot \vec{S}_{j}
\end{equation}
with summation over all the nearest-neighbor bonds, is 
simulated with the SW-type cluster algorithm in which 
all the spin clusters are flipped with $50 \%$ probability 
at each Monte Carlo step (MCS). Although the original 
SW algorithm~\cite{SW} can only be applied to the 
Potts model~\cite{PottsRev}, vector spin models such as the 
Heisenberg model can be treated by constructing spin clusters 
with respect to the Ising element of vector spins projected onto 
a randomly-chosen direction at each MCS~\cite{Wolff}. 

At the critical point $T_{\rm c}$, all the physical quantities can be 
treated with the NER scheme. However, in the off-critical region, 
situation changes drastically. The spontaneous magnetization is 
vanishing above $T_{\rm c}$ and its temperature dependence 
can only be analyzed for $T \leq T_{\rm c}$. Although the absolute 
value of it shows a diverging behavior for $T>T_{\rm c}$, such 
a behavior is nothing but that of the square root of the magnetic 
susceptibility. While the magnetic susceptibility shows diverging 
behaviors in the both sides of $T_{\rm c}$, such a behavior is 
observed after subtracting the contribution from the spontaneous 
magnetization for $T<T_{\rm c}$. Critical exponents of the susceptibility 
and magnetization are different, and NER analysis of a quantity including 
two critical exponents is quite complicated. Moreover, discontinuity 
of relaxation behaviors below and above $T_{\rm c}$ results in the 
restriction of initial states in the NER process. That is, NER started 
from the perfectly-ordered state (corresponding to the configuration 
at $T=0$) can only be applied for $T \le T_{\rm c}$, and that from the 
perfectly-disordered states (one of the configurations at $T=\infty$) 
for $T \ge T_{\rm c}$. 

To summarize the above arguments, the spontaneous magnetization 
can be analyzed from the perfectly-ordered state for $T \le T_{\rm c}$, 
and the magnetic susceptibility from the perfectly-disordered states 
for $T \ge T_{\rm c}$. Although other physical quantities can also be 
treated in principle, those derived from the temperature derivative 
(i.e.\ correlation with energy, e.g.\ the specific heat) show larger 
fluctuations, and the correlation length is evaluated indirectly 
(from the scale dependence of the correlation function or from 
the wave-number dependence of the magnetic susceptibility), 
and therefore they are not preferred for precise estimation. 
The scaled critical exponents $\beta/\nu$ and $\gamma/\nu$ 
can be evaluated from the NE-ES, and the bare exponent 
$\gamma$ from the temperature scaling of the magnetic 
susceptibility as will be seen later. All the critical exponents 
can be obtained from these three exponents through the 
scaling relations. Although the bare exponent $\beta$ can 
also be estimated from the temperature scaling of the absolute 
value of magnetization, it is not as accurate as $\gamma$. 
Details will be explained in the Appendix.

Next, established scaling formulas are briefly reviewed. 
The DFSS for a quantity $Q$ is expressed as 
\begin{equation}
Q(t,L;T) \sim L^{x_{Q}/\nu}f[L/\xi(T),t/\tau(T)],
\end{equation}
with the simulation time $t$, linear size $L$, 
critical exponent $x_{Q}$ defined in 
$Q(\infty,\infty;T) \sim (T-T_{\rm c})^{-x_{Q}}$ 
for $T \to T_{\rm c}$, scaling function $f$, 
correlation length $\xi(T) \sim (T-T_{\rm c})^{-\nu}$, 
and correlation time  $\tau(T) \sim (T-T_{\rm c})^{-z\nu}$ 
in local-update algorithms. 
Assuming equivalence of the functional form of $f$ 
with respect to $t$ and $L$, these two parameters 
are related with each other as $L \sim t^{1/z}$, or 
\begin{equation}
\label{DFSSeq}
Q(t,T) \sim t^{x_{Q}/(z\nu)}f[t^{1/(zv)}(T-T_{\rm c})]
\end{equation}
for a fixed system size. From this scaling form, 
the critical point $T_{\rm c}$ can be evaluated 
from the power-law simulation-time dependence 
of $Q(t,T_{\rm c})$, and an off-critical scaling 
$t^{-x_{Q}/(z\nu)}Q(t,T)$ vs.\ $t^{1/(z\nu)}(T-T_{\rm c})$ 
is derived.

Such a formula does not hold in cluster algorithms, 
because the stretched-exponential critical relaxation 
is not consistent with the power-law size dependence. 
Then, the NE-ES is derived from the critical simulation-time 
dependence, $Q(t;T_{\rm c}) \sim \exp (ct^{\sigma})$ 
(in the NER from the perfectly-disordered states), 
and the equilibrium size dependence at $T_{\rm c}$, 
$Q(L;T_{\rm c}) \sim L^{x_{Q}/\nu}$. 
Combining these formulas, we have 
$L^{-x_{Q}/\nu}Q(t,L;T_{\rm c}) \sim \exp(ct^{\sigma}-\ln L^{x_{Q}/\nu})$, 
or in a more general form corresponding to Eq.~(\ref{DFSSeq}), 
\begin{equation}
Q(t,L;T_{\rm c})
\sim L^{x_{Q}/\nu} f_{\rm sc}(ct^{\sigma}-\ln L^{x_{Q}/\nu}),
\end{equation}
with a scaling function $f_{\rm sc}$ on the NE-ES. 
This scaling form has been confirmed in classical spin 
systems~\cite{Nonomura14,Nonomura16} and in a 
quantum phase transition~\cite{Nonomura20}.

\section{Temperature scaling}
Similarly to the NE-ES, the temperature scaling 
in cluster algorithms is derived from the onset and 
equilibrium behaviors. Namely, from the initial-time 
critical relaxation  $Q(t;T_{\rm c}) \sim \exp (ct^{\sigma})$ 
and the temperature dependence in equilibrium 
$Q(\infty,T) \sim (T-T_{\rm c})^{-x_{Q}}$, we have 
$Q(t,T) (T-T_{\rm c})^{x_{Q}} \sim \exp[ct^{\sigma}+\ln (T-T_{\rm c})^{x_{Q}}]$, or 
\begin{equation}
\label{tsccl}
Q(t,T)
\sim (T-T_{\rm c})^{-x_{Q}}f_{\rm tsc}[ct^{\sigma}+\ln (T-T_{\rm c})^{x_{Q}}],
\end{equation}
with a scaling function $f_{\rm tsc}$ on the temperature scaling. 
Although the above derivation seems more nontrivial than that of 
the NE-ES, usage of the initial-time critical-relaxation formula can 
be justified in comparison with the off-critical scaling (\ref{DFSSeq}), 
which consists of the initial-time dependence at $T_{\rm c}$ and its 
modification by a scaling function with temperature dependence. 

The above derivation is also possible in local-update 
algorithms. From the initial-time critical relaxation 
$Q(t,T_{\rm c}) \sim t^{x_{Q}/(z \nu)}$ and 
the temperature dependence in equilibrium 
$Q(\infty,T) \sim (T-T_{\rm c})^{-x_{Q}}$, we result in 
$Q(t,T)(T-T_{\rm c})^{x_{Q}} \sim [t^{1/(z \nu)}(T-T_{\rm c})]^{x_{Q}}$, or 
\begin{equation}
\label{tsclu}
Q(t,T) \sim (T-T_{\rm c})^{-x_{Q}}f_{\rm tsc}[t^{1/(zv)}(T-T_{\rm c})].
\end{equation}
In comparison with the conventional off-critical scaling form 
(\ref{DFSSeq}), the prefactor of the scaling function is 
changed from $t^{x_{Q}/(z \nu)}$ to $(T-T_{\rm c})^{-x_{Q}}$ 
in the present formalism. 

\section{Numerical results}
\subsection{Swendsen-Wang algorithm}
\begin{figure}
\includegraphics[width=88mm]{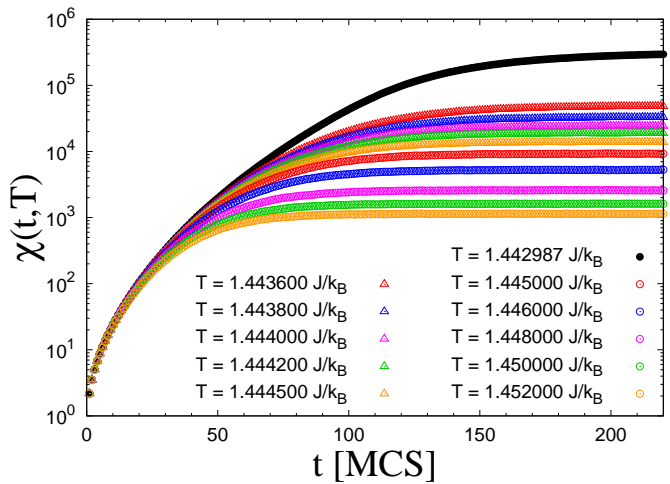}
\caption{Simulation-time dependence of the magnetic susceptibility 
for $L=560$ at $T_{\rm c}$ and various temperatures used for 
the temperature scaling in the SW algorithm. The susceptibility 
decreases monotonically as the temperature increases.
}
\label{figorgcl}
\end{figure}
First, we verify the temperature scaling in cluster algorithms (\ref{tsccl}) 
with the Swendsen-Wang (SW) algorithm. Here we concentrate 
on the magnetic susceptibility, i.e.\ $Q(t,T) \to \chi(t,T)$ 
and $x_{Q} \to \gamma$ in Eq.~(\ref{tsccl}). 
In our previous article to investigate the 3D Heisenberg model with the 
NE-ES based on the SW algorithm~\cite{Nonomura16}, the maximum 
system size was $L=560$. Here we also take $L=560$ and $225$ 
Monte Carlo steps (MCS), and average $4 \times 10^{4}$ 
random-number sequences (RNS). The raw data for 
various temperatures (from $T=1.4436 J/k_{\rm B}$ to 
$1.4520 J/k_{\rm B}$) are shown in Fig.~\ref{figorgcl}, together 
with the data at the most probable value of the critical point, 
$T_{\rm c}=1.442987 J/k_{\rm B}$~\cite{Nonomura16}. 
At $t=225$MCS, $\chi$ for $T=1.4436 J/k_{\rm B}$ 
is about $1/6$ of that at $T=T_{\rm c}$, while that 
at $T=1.4520 J/k_{\rm B}$ is about $1/40$ of that 
at $T=1.4436 J/k_{\rm B}$. Although the range of 
temperature for scaling does not seem so wide, 
that of $\chi$ is actually wide enough. 
In general, the temperature range of scaling is determined 
by the system size in the vicinity of $T_{\rm c}$, and by 
the temperature itself far from $T_{\rm c}$. Although the 
present formulation is based on the diverging behavior 
$\chi(t=\infty,L=\infty,T) \sim (T-T_{\rm c})^{-\gamma}$ 
for $T \to T_{\rm c}$, the actual finite-size behavior is 
saturated with $\chi(t=\infty,L,T_{\rm c}) \sim L^{\gamma/\nu}$, 
and the range of scaling near $T_{\rm c}$ increases as $L$ 
increases. On the other hand, as temperature becomes 
away from $T_{\rm c}$, the weight of the correction 
terms to scaling increases independently of $L$. 

\begin{figure}
\includegraphics[width=88mm]{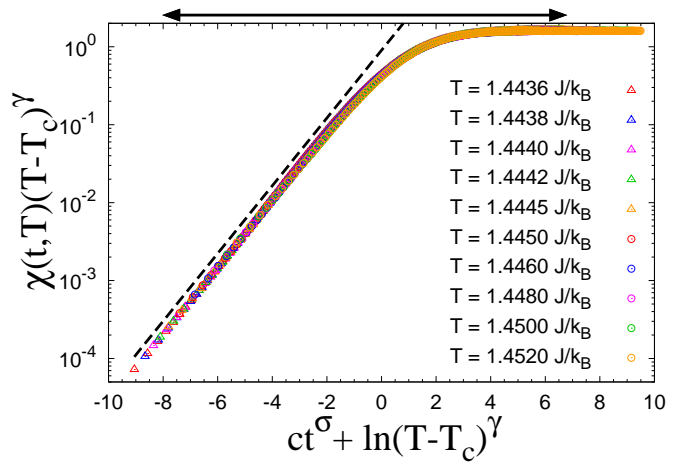}
\caption{Temperature scaling plot of the data in Fig.~\ref{figorgcl} using 
$T_{\rm c}=1.442987 J/k_{\rm B}$ and $\sigma=0.47$~\cite{Nonomura16} 
with $\gamma=1.3945(19)$ and $c=1.2595(43)$ in a semi-log scale. 
The arrow specifies the range of data used for the fitting, and the dashed 
line corresponds to a simple exponential curve as a guide for eyes. 
Here the data for $t=1$ MCS are not included.
}
\label{figtsccl}
\end{figure}
These data are scaled with Eq.~(\ref{tsccl}) in Fig.~\ref{figtsccl}, 
namely the scaling plot of $\chi(t,T)(T-T_{\rm c})^{\gamma}$ versus 
$ct^{\sigma}+\ln (T-T_{\rm c})^{\gamma}$ in a semi-log scale 
using $T_{\rm c}=1.442987(2) J/k_{\rm B}$ and $\sigma=0.47(1)$ 
evaluated in Ref.~\cite{Nonomura16}. 
Since we only take the data rather far away from $T_{\rm c}$, 
precise evaluation of $T_{\rm c}$ is difficult within the present 
scheme. It is also the case in the relaxation exponent $\sigma$. 
This exponent is characteristic to the critical relaxation in cluster 
algorithms, and appearance of it in Eq.~(\ref{tsccl}) is just a trace 
of behaviors at $T_{\rm c}$. Then, it should be determined from 
the critical-relaxation data, not from the off-critical ones. 
The fitting parameters $\gamma$ and $c$ 
are estimated by minimizing the mutual residuals of these data. 
Although every two sets of the data can be scaled with each other, they 
are not independent and error bars cannot be evaluated in a simple way. 
Then, we average the mutual residuals between the nearest-neighbor 
temperatures, determine the range of fitting by minimizing the averaged 
residual as shown by arrows in Fig.~\ref{figtsccl}, and obtain 
\begin{equation}
\label{SWgamma}
\gamma=1.3945 \pm 0.0019,\ c=1.2595 \pm 0.0043.
\end{equation}
Combining this estimate with $\gamma/\nu=1.972 \pm 0.007$ obtained 
from the NE-ES at $T_{\rm c}$~\cite{Nonomura16}, we have 
\begin{equation}
\label{est-nu}
\nu=0.707 \pm 0.003.
\end{equation}

\subsection{Metropolis algorithm}
\begin{figure}
\includegraphics[width=88mm]{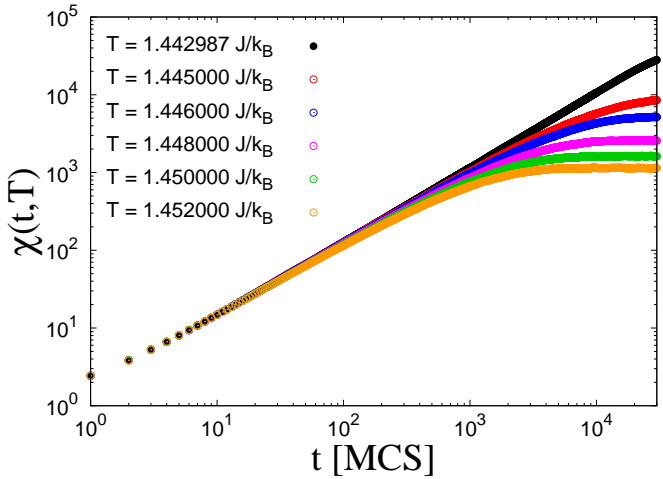}
\caption{Simulation-time dependence of the magnetic susceptibility for 
$L=200$ at $T_{\rm c}$~\cite{Nonomura16} and various temperatures 
used for the temperature scaling in the Metropolis algorithm. 
}
\label{figorglu}
\end{figure}
Next, we testify the temperature scaling in local-update algorithms 
(\ref{tsclu}) based on the Metropolis algorithm, and compare it with 
the standard off-critical scaling (\ref{DFSSeq}) for the same data. 
Here we also consider the magnetic susceptibility and take 
$Q(t,T) \to \chi(t,T)$ and $x_{Q} \to \gamma$ in these formulas. 
We take $L=200$ and $3 \times 10^{4}$ MCS, and average 
$2 \times 10^{4}$ RNS. The raw data at $T_{\rm c}$~\cite{Nonomura16} 
and for various temparatures (from $T=1.445 J/k_{\rm B}$ to 
$1.452 J/k_{\rm B}$) in a log-log scale in Fig.~\ref{figorglu}. 
Since the power-law relaxation at $T_{\rm c}$ is much slower than 
the stretched-exponential critical relaxation in the SW algorithm, 
much longer MCS are required and therefore the system size is 
reduced. The data at $T_{\rm c}$ still show a power-law behavior 
at $t=3 \times 10^{4}$. When we attempt to evaluate $T_{\rm c}$ 
with the conventional NER, relaxation data at $T=1.443 J/k_{\rm B}$ 
cannot be distinguished from the present data at $T_{\rm c}$, and 
the resolution of $T_{\rm c}$ becomes of one order lower than the 
one in Ref.~\cite{Nonomura16}. In comparison with the previous 
subsection, the lowest temperature for scaling is increased 
in response to reduction of the system size, and the highest 
one is the same.

\begin{figure}
\includegraphics[width=88mm]{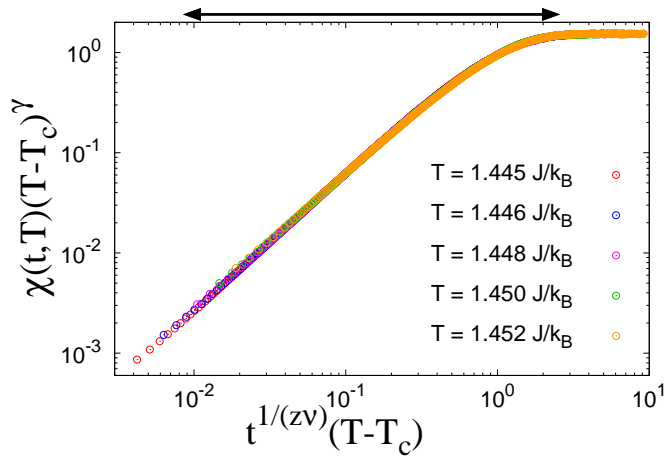}
\caption{Temperature scaling plot of the data in Fig.~\ref{figorglu} 
using $T_{\rm c}=1.442987 J/k_{\rm B}$~\cite{Nonomura16} with 
$\gamma=1.4039(32)$ and $z \nu=1.4866(64)$ in a log-log scale. 
The arrow specifies the range of data used for the fitting. 
Here the data for $t=1,2$ MCS are not included.
}
\label{figtsclu}
\end{figure}
These data are scaled with Eq.~(\ref{tsclu}) in Fig.~\ref{figtsclu}, 
namely the scaling plot of $\chi(t,T)(T-T_{\rm c})^{\gamma}$ 
versus $t^{1/(z \nu)}(T-T_{\rm c})$ in a log-log scale using 
$T_{\rm c}=1.442987(2) J/k_{\rm B}$~\cite{Nonomura16}. 
The fitting parameters $\gamma$ and $z \nu$ are estimated 
by minimizing the mutual residuals of these data. Since the 
relaxation process is much slower than that in the previous 
subsection, the number of data is further increased. 
When all the data are scaled with an equal weight, the 
contribution in the vicinity of equilibrium becomes dominant 
and the functional form in the whole simulation-time regions 
cannot be reproduced anymore. Then, we reduce the density 
of data as sparse as that for $51 \sim 100$ MCS in 
a log scale by averaging the sequential data points. 
That is, we take $100$ points for $1 \sim 100$ MCS, 
$50$ points for $101 \sim 200$ MCS, $60$ points for 
$201 \sim 500$ MCS, $50$ points for $501 \sim 1,000$ 
MCS, $50$ points for $1,001 \sim 2,000$ MCS, $60$ points 
for $2,001 \sim 5,000$ MCS, $50$ points for $5,001 \sim 10,000$ 
MCS, $50$ points for $10,001 \sim 20,000$ MCS, and $20$ points 
for $20,001 \sim 30,000$ MCS; totally we take $490$ points for 
$1 \sim 30,000$ MCS for the fitting. Based on these set of data and 
the fitting scheme similarly to that in the previous subsection, we have 
\begin{equation}
\label{MPgamma}
\gamma=1.4039 \pm 0.0032,\ z\nu=1.4866 \pm 0.0064.
\end{equation}
Combining this estimate with $\nu$ in Eq.~(\ref{est-nu}), we arrive at
\begin{equation}
\label{dynce}
z=2.10 \pm 0.01.
\end{equation}

\begin{figure}
\includegraphics[width=88mm]{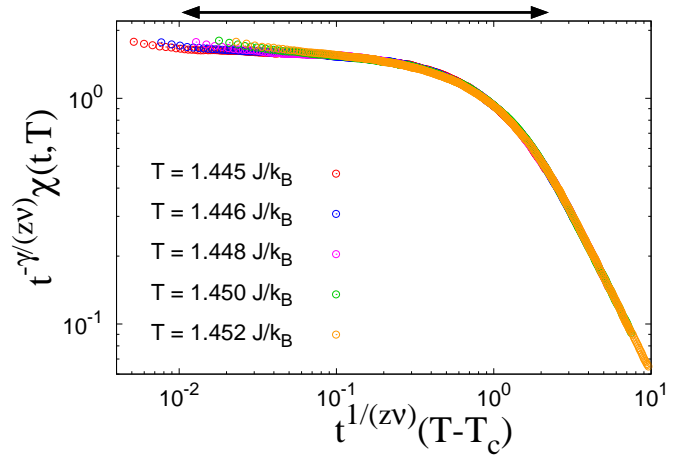}
\caption{Off-critical scaling plot of the data in Fig.~\ref{figorglu} based on 
the DFSS using $T_{\rm c}=1.442987 J/k_{\rm B}$~\cite{Nonomura16} 
with $\gamma=1.4024(57)$ and $z \nu=1.4773(67)$ in a log-log scale. 
The arrow specifies the range of data used for the fitting. 
Here the data for $t=1 \sim 3$ MCS are not included.
}
\label{figdfsslu}
\end{figure}
Finally, we analyze the same data (those in Fig.~\ref{figorglu} 
after the above thinning-out process) with the 
standard off-critical scaling~(\ref{DFSSeq}), namely 
the scaling plot of $t^{-\gamma/(z \nu)}\chi(t,T)$ versus 
$t^{1/(z \nu)}(T-T_{\rm c})$ as shown in Fig.~\ref{figdfsslu}. 
Using $T_{\rm c}=1.442987(2) J/k_{\rm B}$~\cite{Nonomura16} 
and the above fitting scheme, we have 
\begin{equation}
\label{MPconv}
\gamma=1.4024 \pm 0.0057,\ z\nu=1.4773 \pm 0.0067.
\end{equation}
Combining this estimate with $\nu$ in Eq.~(\ref{est-nu}), we obtain 
\begin{equation}
z=2.09 \pm 0.01.
\end{equation}

\section{Discussion}
According to the most precise evaluation of the critical exponents 
of the 3D Heisenberg model until present~\cite{Campostrini02}, 
the exponents treated in the present article were given by 
$\gamma=1.3957(22)$ and $\nu=0.7113(11)$ by equilibrium 
Monte Carlo simulations. Our estimate of $\gamma$ based on 
the SW algorithm (\ref{SWgamma}) is comparable with this one. 
Although ours of $\nu$ (\ref{est-nu}) is rather underestimated, 
it is still within the error bar. Note that this tendency is not due 
to the present analysis, but the one based on the NE-ES at 
$T_{\rm c}$, $\gamma/\nu=1.972(7)$~\cite{Nonomura16}. 
From the estimates in Ref.~\cite{Campostrini02}, it is given 
by $\gamma/\nu=1.962(4)$, and the underestimation of $\nu$ 
simply originates from the overestimation of $\gamma/\nu$. 
Actually, in Ref.~\cite{Campostrini02} the above MC analysis 
was coupled with the high-temperature expansion analysis, and 
they obtained more precise estimates $\gamma=1.3960(9)$ and 
$\nu=0.7112(5)$. Our estimate of $\gamma$ is still consistent 
with it, even though it is rather underestimated.

The tendency of underestimation can be understood from the finite-size 
behavior of physical quantities in the vicinity of equilibrium. As explained 
in the previous section, the temperature scaling is based on the diverging 
behavior of physical quantities, e.g.\ $\chi(T) \sim (T-T_{\rm c})^{-\gamma}$ 
for $T \to T_{\rm c}$. However, such a behavior is only observed 
in the thermodynamic limit, and in finite systems it saturates as 
$\chi(L,T_{\rm c}) \sim L^{\gamma/\nu}$ even at $T=T_{\rm c}$. 
Then, when the data too close to $T_{\rm c}$ in comparison with 
$L$ are taken for the fitting, those become smaller than the ones 
expected from Eq.~(\ref{tsccl}), which results in the underestimation 
of $\gamma$. On the other hand, the data far from $T_{\rm c}$ does 
not converge as sharp as a power with respect to $T-T_{\rm c}$. 
When the data too far away from $T_{\rm c}$ are used for the fitting, 
those become larger than the ones expected from Eq.~(\ref{tsccl}), 
which also causes the underestimation of $\gamma$. 

Our estimate of $\gamma$ based on the temperature scaling 
in the Metropolis algorithm (\ref{MPgamma}) is overestimated 
(it is consistent with the previous estimate within $2\sigma$). 
Although that based on the conventional off-critical scaling 
in the Metropolis algorithm (\ref{MPconv}) is consistent with 
the previous one, it is due to large error bars and the most 
probable value itself is comparable with the one in 
Eq.~(\ref{MPgamma}) and is also overestimated. 
Even in the data in the Metropolis algorithm, tendency of 
underestimation in the vicinity of equilibrium is the same as 
those in the SW algorithm, and this tendency of overestimation 
originates from the early-time nonequilibrium behavior. 
The dynamical critical exponent $z$ is specific to the 
power-law critical relaxation in local-update algorithms, 
and the present estimate (\ref{dynce}) may be comparable 
with that in the 3D Ising model, $z=2.055(10)$~\cite{Ito00}. 
There were no previous studies on the dynamical critical 
behaviors in the 3D Heisenberg model, and we cannot 
argue this slight discrepancy in $z$ too seriously at present. 

Although the temperature scaling holds both in the SW and 
Metropolis algorithms, combination with the SW algorithm 
seems much better in the present analysis. Much larger 
systems can be treated owing to faster relaxation, and 
therefore critical phenomena can be evaluated more 
precisely. Moreover, origin of the discrepancy from 
the previous estimate can be understood naturally. 
In addition, the temperature scaling can be compared 
with the conventional off-critical scaling in the Metropolis 
algorithm. While the two fitting parameters are separated in 
the temperature scaling, they are coupled in the conventional 
off-critical scaling. Then, the error bar becomes twice larger in 
the latter, even though the most probable value of the estimate 
is comparable.

In the present article, we proposed the following procedure to 
determine critical phenomena with the cluster NER scheme:
\begin{enumerate}
\item Determine $T_{\rm c}$ by the NE-ES on the magnetization 
and/or magnetic susceptibility.
\item Determine $\beta/\nu$ and $\gamma/\nu$ by the NE-ES 
together with the above $T_{\rm c}$.
\item Determine $\gamma$ by the temperature scaling using the 
above $T_{\rm c}$.
\item Evaluate other critical exponents through the scaling relations.
\end{enumerate}
This is a minimum procedure, and 
precise evaluation of $\beta$ within the present scheme seems difficult 
at present, as explained in the Appendix. However, from the scaling 
relation $\alpha + 2\beta +\gamma = 2$ and the hyperscaling relation 
$2- \alpha = d\nu$, we have $2\beta/\nu + \gamma/\nu = d$. 
That is, evaluation of $\beta$ is actually not necessary for 
the study on critical phenomena. If the critical exponent 
$\nu$ can be estimated from the temperature scaling of 
the correlation length $\xi$, the universality class can be 
identified only with the present scheme. Nevertheless, 
precise evaluation of $T_{\rm c}$ is not possible within this scheme, and 
the NE-ES of the critical relaxation is indispensable for the cluster NER.

\section{Summary}
In the present article, we proposed a new scaling theory in the 
nonequilibrium relaxation process called as the temperature scaling, 
and we confirmed this theory on the magnetic susceptibility in the 
3D Heisenberg model. When the temperature scaling was combined 
with the Swendsen-Wang (SW) algorithm, it worked very well and our 
estimate of the critical exponent $\gamma=1.3945(19)$ is comparable 
with the previous best estimate. When it was combined with the 
Metropolis algorithm, it worked as well as the conventional off-critical 
scaling, but not as well as the case with the SW algorithm, because 
of limitation of system sizes owing to slow relaxation. 

\begin{acknowledgments}
The present study was supported by JSPS (Japan) KAKENHI Grant 
No.~20K03777. 
The random-number generator MT19937~\cite{MT} was used for 
numerical calculations. Part of the calculations were performed on 
the Supercomputer Center at the Institute for Solid State Physics, 
the University of Tokyo, and on the Numerical Materials Simulator 
at the National Institute for Materials Science.
\end{acknowledgments}

\appendix

\section{Magnetization in the SW algorithm}
\begin{figure}
\includegraphics[width=88mm]{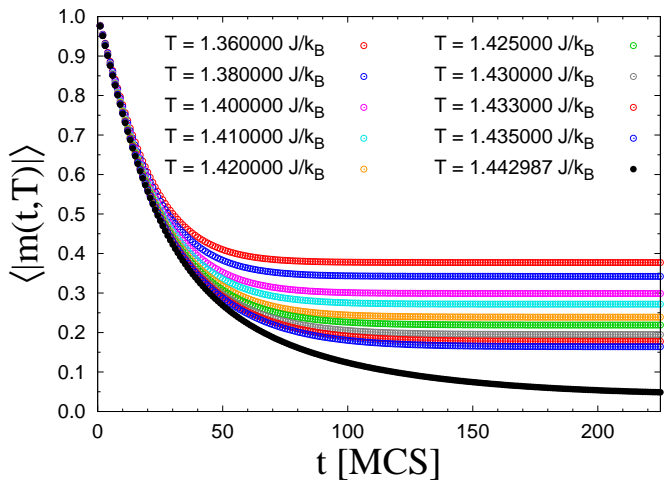}
\caption{Simulation-time dependence of the absolute value 
of magnetization for $L=560$ at $T_{\rm c}$ and various 
temperatures in the SW algorithm. The magnetization 
decreases monotonically as the temperature increases.
}
\label{msorgfig}
\end{figure}
\begin{figure}
\includegraphics[width=88mm]{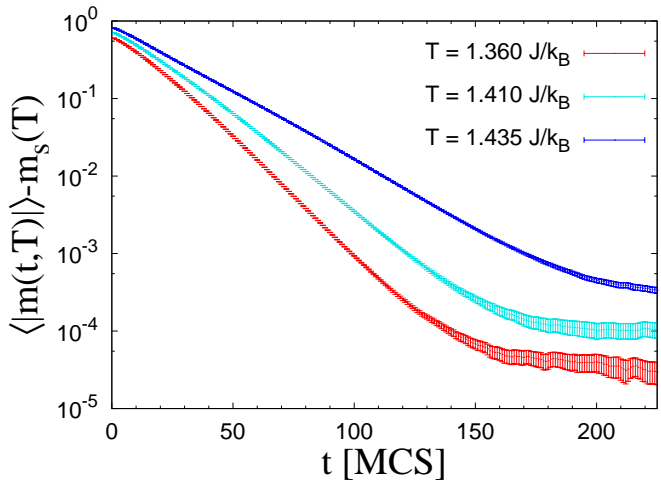}
\caption{
Simulation-time dependence of the decaying part of the absolute value 
of magnetization based on Eq.~(\ref{expdec}) in a semi-log scale at 
$T=1.360 J/k_{\rm B}$, $1.410 J/k_{\rm B}$ and $1.435 J/k_{\rm B}$ 
(from bottom to top). 
}
\label{expdecfig}
\end{figure}
Even if Monte Carlo simulations are started from the perfectly-ordered state, 
the sign of magnetization may change in each step by a global flip of large 
clusters in the relaxation process in cluster algorithms. When the data of 
different random-number sequences are averaged, cancellation of signs 
takes place and the averaged results become meaningless. 
Then, in the cluster NER, we take the absolute value of magnetization. 
Here we start from the perfectly-ordered state, simulate the $L=560$ 
system during $225$ MCS with the SW algorithm, and average 
$4 \times 10^{4}$ RNS. The relaxation data for various temperatures 
(from $T=1.360 J/k_{\rm B}$ to $1.435 J/k_{\rm B}$ and at $T_{\rm c}$) 
are displayed in Fig.~\ref{msorgfig}. 

Although the data at $T_{\rm c}$ decay on a stretched-exponential 
curve and do not arrive at equilibrium at $t=225$ MCS, other data 
for $T<T_{\rm c}$ seem to be already in equilibrium at that 
simulation time. Such relaxation behaviors are described 
by the following formula, 
\begin{equation}
\label{expdec}
\langle |m(t,T)| \rangle = m_{\rm s}(T)+A(T)\exp(-C(T)t), 
\end{equation}
with the spontaneous magnetization $m_{\rm s}(T)$ and fitting 
parameters $A(T)$ and $C(T)$. This formula was confirmed in 
the 2D Ising model in the Wolff algorithm~\cite{Nonomura14}, 
while the stretched-exponential relaxation was reported 
in the local-update algorithms~\cite{Ito93,Stauffer96}. 
This relaxation formula is verified in Fig.~\ref{expdecfig} 
by fitting the data with Eq.~(\ref{expdec}) and plotting 
$\langle |m(t,T)| \rangle - m_{\rm s}(T)$ versus $t$ 
in a semi-log scale at $T=1.360 J/k_{\rm B}$, $1.410 J/k_{\rm B}$ 
and $1.435 J/k_{\rm B}$ (from bottom to top). Linearity of the data 
reveals validity of Eq.~(\ref{expdec}), and variance of the initial value 
and slope of the data represents explicit temperature dependence of 
the parameters $A(T)$ and $C(T)$ in Eq.~(\ref{expdec}), respectively. 
Such nontrivial $T$-dependence other than that of $m_{\rm s}(T)$ 
makes a scaling analysis based on Eq.~(\ref{expdec}) difficult. 

\begin{figure}
\includegraphics[width=88mm]{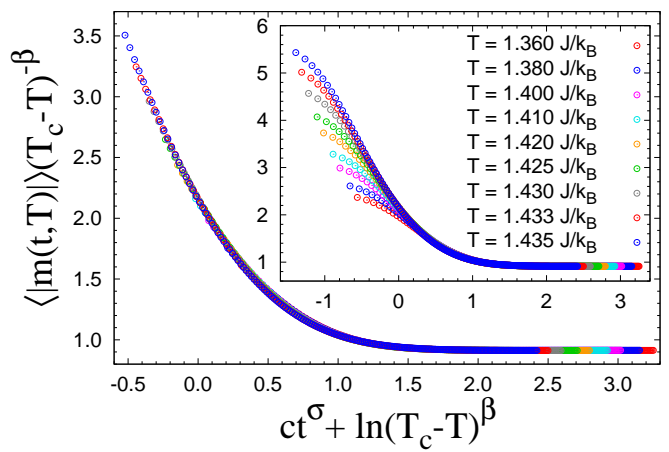}
\caption{Temperature scaling plot of the data in Fig.~\ref{msorgfig} 
after $16$ MCS using $T_{\rm c}=1.442987 J/k_{\rm B}$ and 
$\sigma=0.47$~\cite{Nonomura16} with $\beta=0.3553(10)$ 
and $c=0.3244(3)$. The plot including the relaxation data 
for the initial $15$ MCS is shown in the inset.
}
\label{mstscfig}
\end{figure}
Nevertheless, the temperature scaling still holds 
on this quantity. From the stretched-exponential 
critical relaxation from the perfectly-ordered state, 
$\langle |m(t,T)| \rangle \sim \exp(-c t^{\sigma})$, 
and the temperature dependence in equilibrium, 
$\langle |m(t=\infty,T)| \rangle \sim (T_{\rm c}-T)^{\beta}$, we have 
\begin{equation}
\label{mstsc}
\langle |m(t,T)| \rangle \sim
(T_{\rm c}-T)^{\beta} f_{\rm tsc}[c t^{\sigma} + \ln (T_{\rm c}-T)^{\beta}].
\end{equation}
The data in Fig.~\ref{msorgfig} are scaled with Eq.~(\ref{mstsc}) in 
Fig.~\ref{mstscfig}. Although the initial-time data are not scaled well 
owing to the discrepancy with the exponential decay (\ref{expdec}) 
as shown in the inset of Fig.~\ref{mstscfig}, the scaling formula 
(\ref{mstsc}) actually holds very well for the data from $16$ MCS 
(in the main panel of Fig.~\ref{mstscfig}). 
Similarly to the temperature scaling of the magnetic susceptibility, 
we minimize the mutual residuals of these data 
using $T_{\rm c}=1.442987(2) J/k_{\rm B}$ 
and $\sigma=0.47(1)$~\cite{Nonomura16}. 
We find that the averaged residuals are minimized 
when all the data in the main panel of Fig.~\ref{mstscfig} 
are used for the fitting, and we have 
\begin{equation}
\label{mstscexp}
\beta=0.3553 \pm 0.0010,\ c=0.3244 \pm 0.0003.
\end{equation}
Although the error bars seem small enough, this estimate is 
not consistent with the most precise estimate until present, 
$\beta=0.3689(3)$~\cite{Campostrini02}. 

The background of this discrepancy can be explained by the 
evaluation of $\beta$ from the temperature dependence of 
$m_{\rm s}(T)$ in Eq.~(\ref{expdec}). Up to the leading term, 
it is given by $m_{\rm s}(T)=B_{1}(T_{\rm c}-T)^{\beta}$, and 
using all the data for $T<T_{\rm c}$ in Fig.~\ref{msorgfig}, 
we have $\beta=0.3574(2)$. This estimate is not so different 
from that in Eq.~(\ref{mstscexp}), and not consistent with the 
one in Ref.~\cite{Campostrini02}, neither. On the other hand, 
when we take the next-order term into account as 
\begin{equation}
m_{\rm s}(T)=B_{1}(T_{\rm c}-T)^{\beta}+B_{2}(T_{\rm c}-T)^{2\beta}, 
\end{equation}
we obtain
\begin{eqnarray}
&&\beta = 0.3691 \pm 0.0010,\\
&& B_{1} = 0.988 \pm 0.005,\ B_{2} = -0.107 \pm 0.007.
\end{eqnarray}
This estimate is consistent with the one in Ref.~\cite{Campostrini02}, 
and the coefficient of the next-order term is about $10\%$ of that 
of the leading term. These results tell that the next-order term 
is crucial for the description of the critical phenomena in the 3D 
Heisenberg model based on the temperature dependence of 
the magnetization, and that the temperature-scaling formalism 
based only on the leading term of the temperature dependence 
of physical quantities is not suitable for the magnetization, at least 
in the present model. This mechanism is independent of the update 
algorithms, and therefore we do not consider the Metropolis algorithm here.

\end{document}